\documentclass[prl,twocolumn]{revtex4}
\usepackage{amssymb}
\usepackage{amsmath}
\usepackage{bbm}
\usepackage{graphicx}

\begin{document}

\title{A cluster algorithm for resistively shunted Josephson junctions}
\author{Philipp Werner}
\author{Matthias Troyer}
\affiliation{Institut f{\"u}r theoretische Physik, ETH H{\"o}nggerberg, CH-8093 Z{\"u}rich, Switzerland}

\date{\today}

\hyphenation{si-mi-lar theo-retical}

\begin{abstract}
We present a cluster algorithm for resistively shunted Josephson junctions or similar physical systems, which dramatically improves sampling efficiency. The algorithm combines local updates in Fourier space with rejection-free cluster updates which exploit the symmetries of the Josephson coupling energy. As an application, we consider the superconductor-to-insulator transition in a single junction at intermediate Josephson coupling and determine the temperature dependence of the zero bias resistance as a function of dissipation strength.
\end{abstract}


\maketitle

Resistively shunted Josephson junctions (JJ) at zero temperature undergo a superconductor-to-insulator transition if the shunt resistance $R_s$ is increased beyond a critical value, which equals the quantum of resistance $R_Q = h/4e^2$. This dissipative phase transition was first predicted by Chakravarty \cite{Chakravarty} and Schmid \cite{Schmid} and subsequently studied by several authors \cite{Bulgadaev, Guinea, Fisher&Zwerger, Schoen}. Similar transitions occur in JJ arrays and rich phase diagrams have been worked out using variational approximations \cite{Chakravarty&Ingold1}, mean field theory \cite{Panukov&Zaikin} and semi-classical calculations \cite{Korshunov1}. Recently, experimental realizations of resistively shunted single junctions \cite{Yagi} and JJ-chains \cite{Miyazaki} have allowed to test some of the theoretical predictions. These experimental advances and the rising interest in questions related to dissipation in macroscopic quantum devices have led to new theoretical investigations. Floating phases in linear arrays \cite{Tewari} and complicated phase diagrams for a two-junction system with charge relaxation \cite{Refael1} have been predicted. An extension of the latter work to chains of JJ \cite{Refael2} attempts to explain the intriguing experimental results on the superconductor-to-normal transition in nanowires \cite{Bezryadin}.

It would be desirable to test these sometimes controversial theoretical predictions numerically. Path-integral Monte Carlo studies of a single, resistively shunted JJ have recently been carried out \cite{Herrero&Zaikin, Kimura&Kato}. However, the lack of efficient algorithms has so far proven to be an obstacle and the results in Ref. \cite{Kimura&Kato} even disagree with analytical predictions. In this Letter, we present a new cluster algorithm which dramatically improves sampling efficiency compared to local update schemes and allows to simulate the junction at more than 100 times lower temperature. Such efficient updates for single junctions will be important for the simulation of JJ chains and arrays.   

Dissipation in the shunt resistor is introduced by coupling the phase difference $\phi$ across the junction linearly to an infinite set of harmonic oscillators \cite{Caldeira}. For an Ohmic heat bath one finds the effective action \cite{Schoen} (setting $\hbar=1$)
\begin{eqnarray}
S[\phi] &=& \int_0^\beta d\tau \left[\frac{1}{16E_C}\Big(\frac{d\phi}{d\tau}\Big)^2-E_J\cos(\phi)\right]\nonumber\\
&+& \frac{1}{8\pi^2}\frac{R_Q}{R_s}\int_0^\beta \int_0^\beta d\tau d\tau' \frac{(\pi/\beta)^2(\phi(\tau)-\phi(\tau'))^2}{\sin((\pi/\beta)(\tau-\tau'))^2}.\hspace{5mm}
\label{eq1}
\end{eqnarray}  

The first term describes the capacitive coupling between the two superconductors (the effective charging energy of the junction, $E_C=\frac{e^2}{2C}$, sets the energy scale) and the Josephson energy for a coupling strength $E_J$. The last term results from integrating out the heat bath degrees of freedom and introduces dissipation. Because of the Josephson energy and the non-compact nature of the phase variable one cannot directly employ the cluster algorithms available for spin systems \cite{Swendsen&Wang, Wolff, Luijten&Bloete}, as was done in Ref.~\cite{ising}. We therefore propose a new algorithm consisting of two kinds of updates: (i) local updates in Fourier space compatible with the Gaussian terms in (\ref{eq1}) and (ii) rejection-free cluster updates. The fist type of moves assures ergodicity of the algorithm and the second type produces global cluster updates compatible with the energetic constraints from the Josephson term.

We discretize imaginary time into $N$ (assumed odd) time steps $\Delta\tau=\beta/N$ and introduce the dimensionless coefficient $\alpha=R_Q/R_s$. The action (\ref{eq1}) can then be expressed in the simple form 
\begin{equation}
S[\phi] = \sum_{k=0}^{N-1} a_k|\tilde\phi_k|^2-E_J\Delta\tau\sum_{j=0}^{N-1}\cos(\phi_j),
\label{eq3}
\end{equation}  
where $\tilde\phi_k=\sum_{j=0}^{N-1} e^{i\frac{2\pi}{N}jk}\phi_j$ denotes the Fourier transform of $\phi$. The positive coefficients $a_k$ are defined as $a_k= \frac{2}{N}(\tilde g_0-\tilde g_k$), with $g_k$ the Fourier transform of the kernel ($j\ne 0$)
\begin{equation}
g(j) = \frac{1}{32E_C\Delta\tau}(\delta_{j,1}+\delta_{j,N-1})+\frac{\alpha}{8\pi^2}\frac{(\pi/N)^2}{\sin((\pi/N)j)^2}.
\label{eq4}
\end{equation}
Since $\tilde \phi_k = \tilde \phi_{N-k}$, only $\{\tilde\phi_k | k=0,\ldots (N-1)/2\}$ need to be considered. In a local update of the frequency component $\tilde\phi_k$, we choose a new value according to the probability distribution of the Gaussian term in Eq.~(\ref{eq3}), 
\begin{equation}
p(\phi_k) \sim e^{-2a_k|\tilde \phi_k| ^2},
\label{eq5}
\end{equation}
and accept it with probability
\begin{equation}
p(\phi_{\text{old}}\rightarrow \phi_{\text{new}}) =
\min\left(1, e^{-\{S_J[\phi_{\text{new}}]-S_J[\phi_{\text{old}}]\}}\right).
\label{eq6}
\end{equation}
Here $S_J[\phi]=-E_J\Delta\tau\sum_{j=0}^{N-1}\cos(\phi_{j})$ and $\phi_{old}$, $\phi_{new}$ denote the backward Fourier transform of the old and new $k$-space configuration, respectively. Such local updates can be performed in a time $O(N)$ and have recently been used in the simulation of 2D JJ arrays \cite{Capriotti}.

For reasonably large values of $E_J$, local $k$-space updates which introduce phase changes on the order of $2\pi$ will be strongly suppressed, because their sinusoidal shape does not resemble an optimal phase slip path. Algorithms based on local updates alone will therefore be ineffective near the phase transition, where phase slips start to proliferate. A typical path will stay most of the time near one or the other of the minima in the cosine-potential, as shown in Fig.~1. Global moves compatible with this step-like structure would be updates which shift the phases $\phi_j$ by $\pm 2\pi$ in some random interval or reflections about $\phi = n\pi$, $n\in \mathbbm{Z}$, that is the positions of the maxima and minima of the cosine potential.

\begin{figure}[t]
\hfill
\begin{minipage}[b]{\linewidth}
\centering
\includegraphics [angle=0, width=8.5cm]{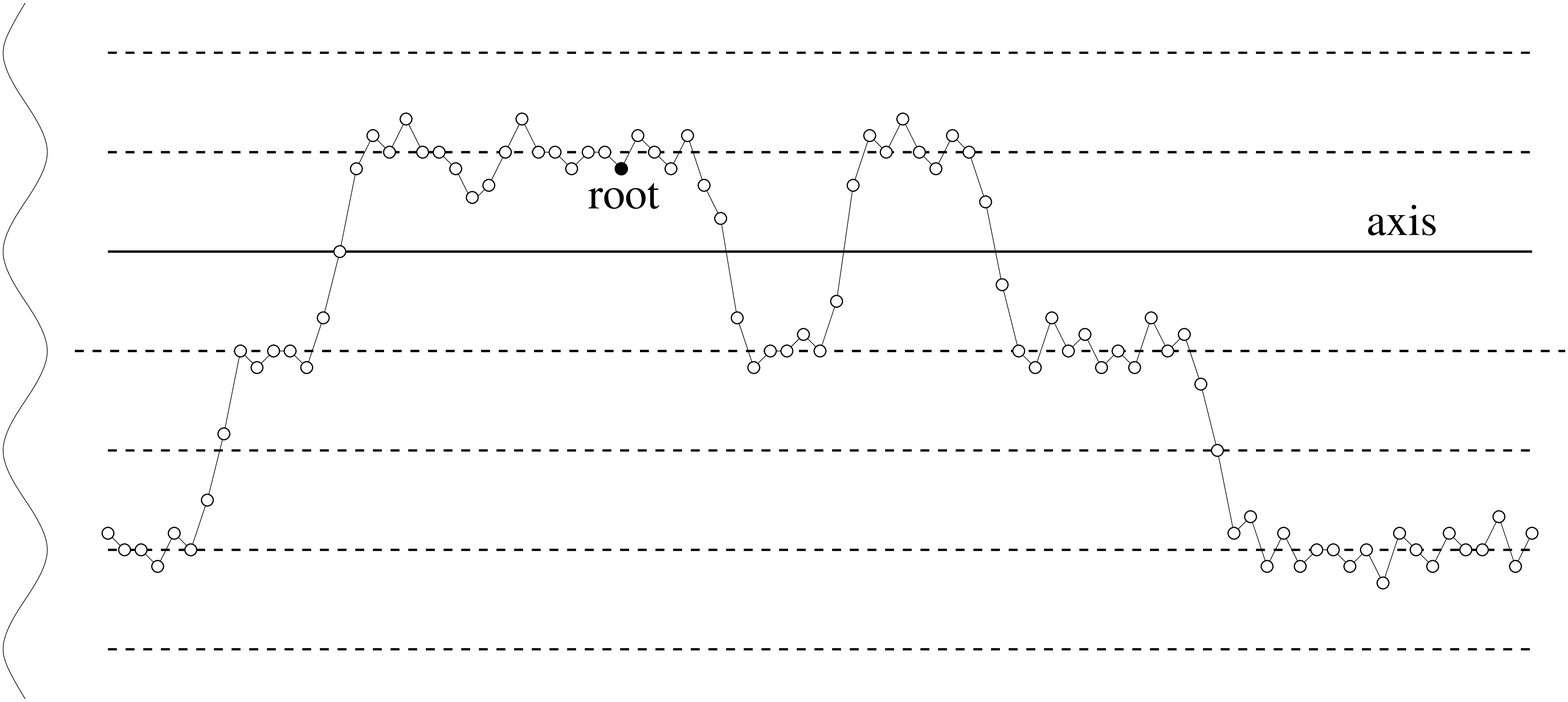}
\end{minipage}
\vspace{1mm}\\
\mbox{}\hfill
\begin{minipage}[b]{\linewidth}
\centering
\includegraphics [angle=0, width=8.5cm]{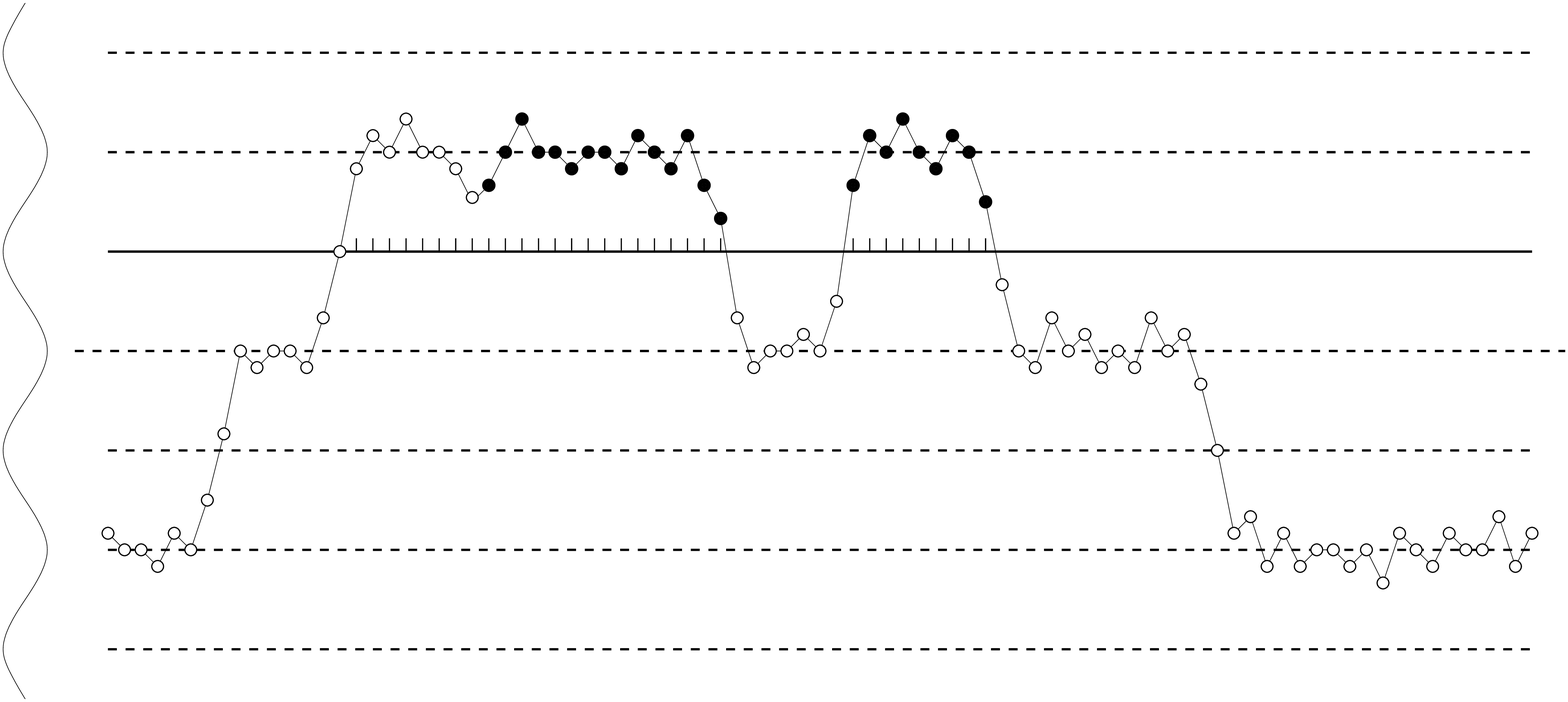}
\end{minipage}
\vspace{1mm}\\
\mbox{}\hfill
\begin{minipage}[b]{\linewidth}
\centering
\includegraphics [angle=0, width=8.5cm]{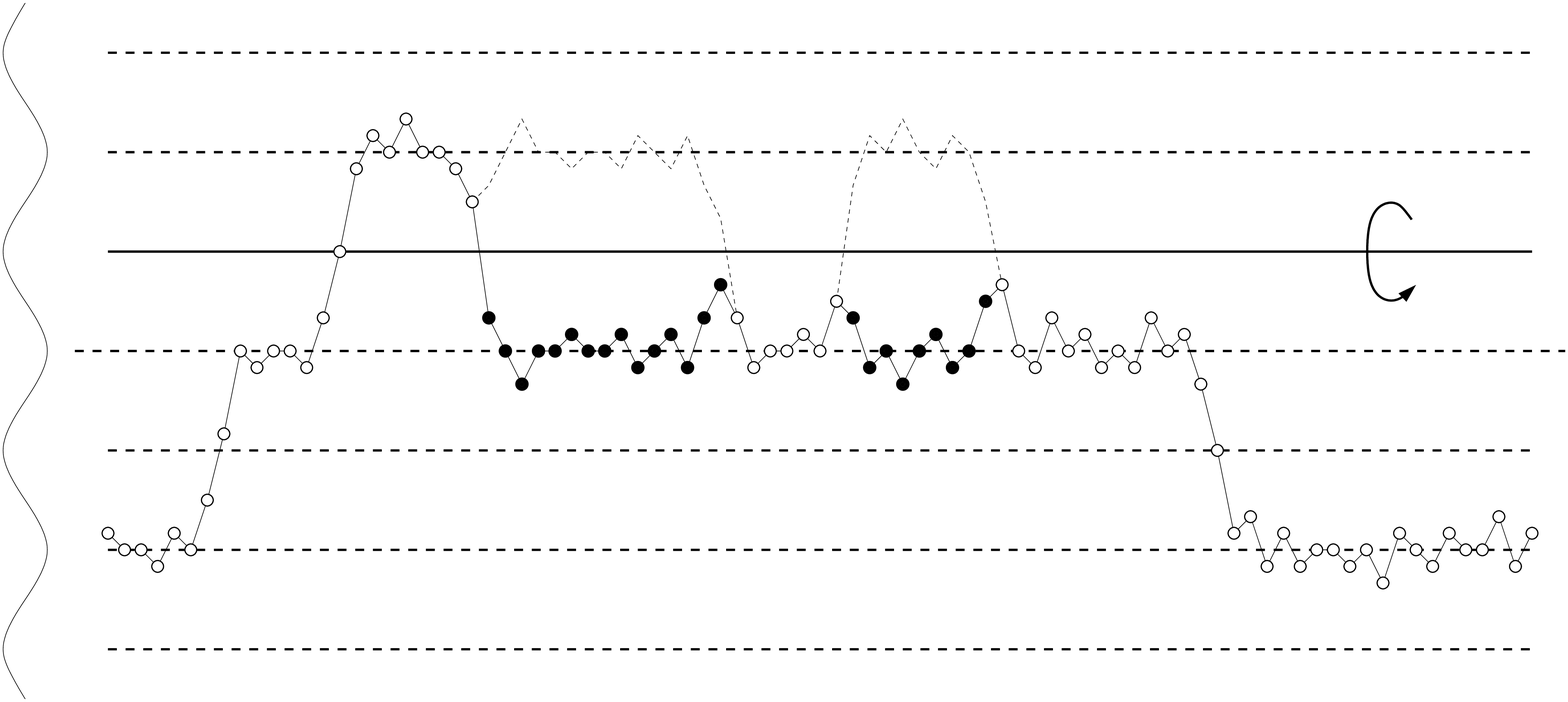}
\end{minipage}
\hfill \mbox{}
\caption{Illustration of the cluster algorithm for non-compact phase variables. The figures show from top to bottom: (i) Original phase configuration. Possible axis positions are indicated with dashed lines, located at the maxima and minima of the cosine potential. The randomly chosen axis and root site of the cluster are marked with the black solid line and black dot respectively. (ii) Cluster of connected sites. The sites which could potentially connect to the root site are indicated with tic marks. (iii) New phase configuration obtained by flipping the cluster around the axis.}
\label{cluster}
\end{figure}

\begin{figure}[t]
\centering
\includegraphics [angle=-90, width= 8.5cm] {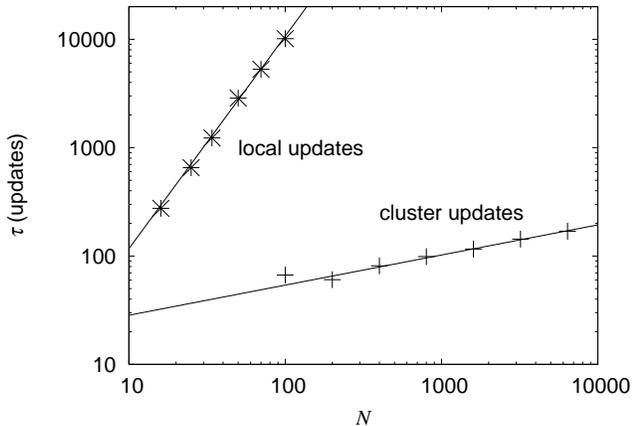}
\caption{Integrated autocorrelation times $\tau$ for $\langle (\phi-\bar\phi)^2\rangle$ versus system size $N$. The data were obtained at the critical point $\alpha=1$ using $E_J/E_C=1$.}
\label{auto}
\end{figure}

We exploit the latter symmetry to design a rejection-free cluster update consisting of the following four steps, which are illustrated in Fig.~\ref{cluster}: (i) An axis $\phi=n^\text{axis}\pi$ with integral $n^\text{axis}$ in the range $[-n_\text{max}, n_\text{max}]$ is randomly chosen (the significance of $n_\text{max}$ is discussed below) and a random site $j_\text{root}$ is picked as the root site of the cluster. We introduce relative coordinates
\begin{equation}
\phi^\text{axis}_j = \phi_j - n^\text{axis}\pi,
\label{eq8}
\end{equation}
which are updated in a cluster move as $\phi_j^\text{axis}\rightarrow -\phi_j^\text{axis}$. Such updates do not change the value of $S_J$. (ii) The cluster of sites connected to $j_\text{root}$ is constructed in a way analogous to other cluster algorithms \cite{Swendsen&Wang, Wolff, Evertz, Krauth, Liu&Luijten}, although the task is complicated by the presence of long ranged interactions. Two sites at positions $i$ and $j$ are connected with probability
\begin{equation}
p(i,j) = \max\Big(0, 1-e^{-\{S[\phi^\text{axis}_i,-\phi^\text{axis}_j]-S[\phi^\text{axis}_i,\phi^\text{axis}_j]\}}\Big),
\label{eq9}
\end{equation}
where
\begin{equation}
S[\phi^\text{axis}_i,-\phi^\text{axis}_j]-S[\phi^\text{axis}_i,\phi^\text{axis}_j] = 8g(i-j)\phi^\text{axis}_i \phi^\text{axis}_j.
\label{10}
\end{equation}
(iii) The phases of the sites $j$ belonging to the cluster are updated according to $\phi_j\rightarrow 2n^\text{axis}\pi-\phi_j$ (in relative coordinates $\phi^\text{axis}\rightarrow -\phi^\text{axis}$). (iv) If necessary, the whole configuration is shifted by multiples of $\pm 2\pi$ in such a way that the mean value $\bar\phi=\frac{1}{N}\sum_{j=0}^{N-1}\phi_j$ is closest to the potential minimum corresponding to $n^\text{axis}=0$. The last step is important to assure detailed balance with a finite $n_\text{max}$. The parameter $n_\text{max}$ must be large enough that the re-centered configuration is contained in the interval $[-n_\text{max}\pi, n_\text{max}\pi]$. This value can be determined during the thermalization of the system.
An alternative to fixing the interval of symmetry points and re-centering the new configurations would be to choose the axis among the $n_\text{max}$ symmetry points above or below $\phi_{j_\text{root}}$.

Using ideas of Luijten and Bl\"ote \cite{Luijten&Bloete}, the cluster move can be performed in a time $O(N\log N)$ despite long ranged interactions, which allows us to compute precise data for large systems (up to $N\approx 10^4$ if $\Delta\tau E_C = 0.25$). On small lattices the data produced by local update schemes are consistent with those obtained using the cluster algorithm, but we find that the results from local update simulations become unreliable for systems larger than $N\approx 100$ because of diverging autocorrelation times. Fig.~\ref{auto} plots the integrated autocorrelation time $\tau$ for $\langle (\phi-\bar\phi)^2 \rangle$ as a function of system size $N$. Even though the CPU time for a cluster update is $O(N\log N)$ as compared to $O(N)$ for a local update, the gain in sampling efficiency is considerable. 

\begin{figure}[t]
\centering
\includegraphics [angle=-90, width= 8.5cm] {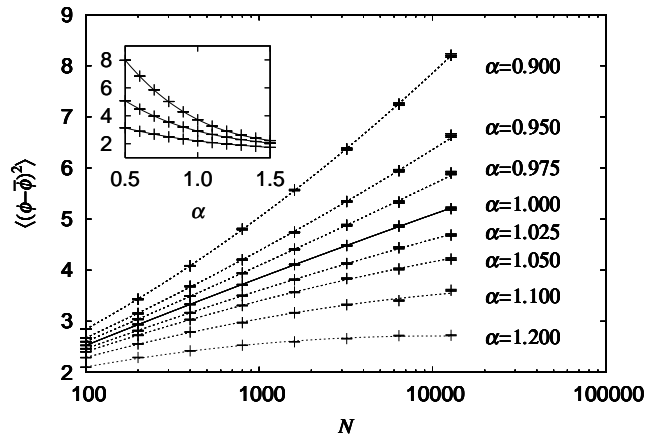}
\caption{$\langle(\phi-\bar\phi)^2\rangle$ plotted as a function of system size (inverse temperature) for several values of dissipation strength $\alpha$. The curves are parabolic fits to the data obtained for $E_J/E_C=1$. The inset shows $\langle(\phi-\bar\phi)^2\rangle$ as a function of $\alpha$ for $E_J/E_C=0.75$ and (from top to bottom) $N=200$, 80 and 36.}
\label{phi2}
\end{figure}

To demonstrate the performance of the new algorithm we study the localization transition at intermediate values of the Josephson coupling energy. Theoretically, this transition is predicted to occur at $\alpha=1$ in the limit of large or small $E_J/E_C$ and it was conjectured that its position is in fact independent of $E_J/E_C$. Local update Monte Carlo simulations have recently been used in Ref.~\cite{Herrero&Zaikin} to investigate the transition at intermediate coupling energy. These authors plotted $\langle(\phi-\bar\phi)^2\rangle$ as a function of dissipation strength $\alpha$ and claimed that the slope of such curves changes abruptly at the critical value $\alpha_c=1$. In the inset of Fig.~\ref{phi2} we show our high accuracy data for the same parameter values as in Ref.~\cite{Herrero&Zaikin}. Even at considerably lower temperatures, $\langle(\phi-\bar\phi)^2\rangle$ changes smoothly around $\alpha=1$, and the method of Ref.~\cite{Herrero&Zaikin} cannot be used to determine the critical point.

Instead, we plot $\langle(\phi-\bar\phi)^2\rangle$ as a function of the inverse temperature $\beta$ (or system size $N$). For $\alpha=1$ one observes an increase proportional to $\log N$. Below $\alpha=1$, the fluctuations grow faster than logarithmically and hence diverge (insulating phase), whereas above $\alpha=1$ the phase fluctuations presumably saturate at some finite value corresponding to a localized configuration (superconductivity). This is illustrated in Fig.~\ref{phi2} for $E_J/E_C = 1$ and a similar result is found for $E_J/E_C = 2$. In the simulations, we used $\Delta\tau E_C=0.25$, so the two coupling strengths correspond to $\Delta\tau E_J=0.25$ and $\Delta\tau E_J=0.5$ respectively, and
the inverse temperature is related to the lattice size by $\beta E_C=N/4$.

\begin{figure}[t]
\centering
\includegraphics [angle=-90, width= 8.5cm] {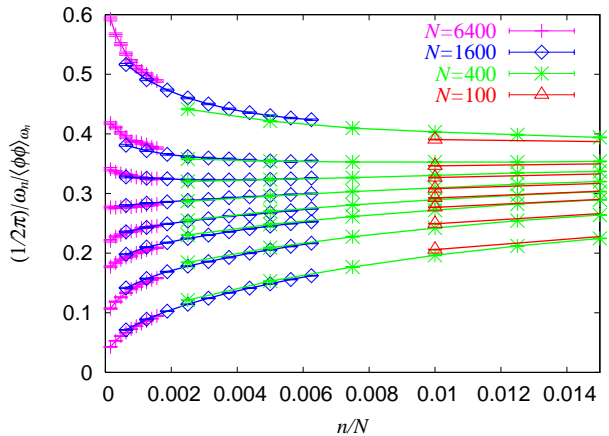}
\caption{$\frac{1}{2\pi}|\omega_n|\langle\phi\phi\rangle_{\omega_n}$ plotted as a function of $n/N$ for different system sizes (inverse temperatures) and dissipation strengths $\alpha$. $E_J/E_C=1$ and the values of $\alpha$ are (from top to bottom): 0.9, 0.95, 0.975, 1, 1.025, 1.05, 1.1 and 1.2.}
\label{matsubara}
\end{figure}

Another method to locate the phase boundary is to look at the Fourier transform $\langle \phi\phi\rangle_{\omega_n}$ of $\langle\phi(0)\phi(\tau)\rangle$ as was done in Refs.~\cite{Schmid, Kimura&Kato}. At low temperature, the Matsubara frequencies $\omega_n=\frac{2\pi n}{\beta}$ are densely spaced and it becomes possible to determine the zero bias resistance $R$ from an extrapolation to $n=0$,
\begin{equation}
\frac{R}{R_Q} = \lim_{n\rightarrow 0} \frac{1}{2\pi}|\omega_n|\langle\phi\phi\rangle_{\omega_n}.
\label{eq11}
\end{equation}   
The critical value of $\alpha$ can be determined from the temperature dependence of $R$. In the $T=0$ superconducting phase, the resistance will decrease to zero with decreasing temperature. If the junction turns insulating, the resistance will increase and saturate at $R_s$, since normal electrons continue to flow through the shunt. In Fig.~\ref{matsubara} we plot $\frac{1}{2\pi}|\omega_n|\langle\phi\phi\rangle_{\omega_n}$ for different temperatures and values of $\alpha$. At small $n/N$, the curves for $\alpha\ge 1.025$ and their extrapolations to $n=0$ decrease with decreasing temperature, while those for $\alpha\le 0.975$ increase.

\begin{figure}[t]
\centering
\includegraphics [angle=-90, width= 8.5cm] {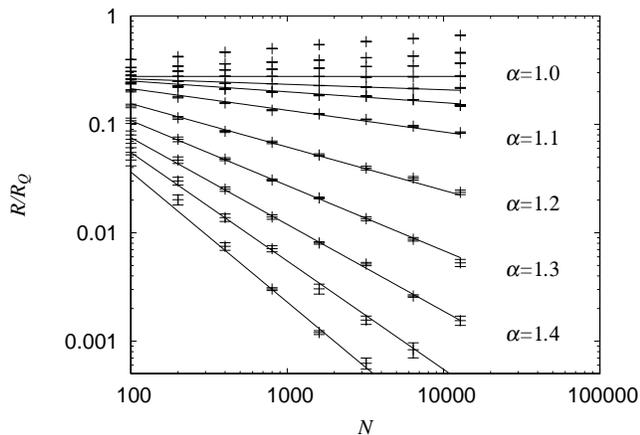}
\caption{Zero bias resistance as a function of temperature for $E_J/E_C=1$ and several values of the dissipation strength. From top to bottom, the data points correspond to $\alpha=0.9$, 0.95, 0.975, 1, 1.025, 1.05, 1.1, 1.2, 1.3, 1.4, 1.5 and 1.6. The lines show a fit of the power-law Eq.~(\ref{eq12}).}
\label{R_beta}
\end{figure}

In order to make the above analysis more precise and quantitative, we performed the extrapolation (\ref{eq11}) using parabolic fits to the first five Matsubara points. The values of $R$ obtained by this procedure are plotted as a function of the inverse temperature in Fig.~\ref{R_beta}. Again, the transition at $\alpha=1$ ($R$ decreasing with inverse temperature above and increasing below the critical value) is clearly visible. Figs.~\ref{matsubara} and \ref{R_beta} show the data for $E_J/E_C=1$ but similar results were obtained for $E_J/E_C=2$. This is the first convincing numerical evidence that the phase transition occurs at $\alpha=1.00(2)$ for intermediate coupling strengths, which supports the conjecture that $\alpha_c$ does not depend on the value of the Josephson coupling.

For $\alpha>1$, Korshunov \cite{Korshunov1} predicted a temperature dependence of the resistance of the form
\begin{equation}
R(T) \sim T^{2\alpha-2}.
\label{eq12}
\end{equation}
In a recent numerical study \cite{Kimura&Kato} a temperature dependence incompatible with Eq.~(\ref{eq12}) was found and the authors even observed a strong dependence of the power-law exponent on the Josephson energy. However, these data were obtained at rather high temperature, where Eq.~(\ref{eq12}) may not be valid.
At the much lower temperatures accessible with the cluster algorithm, we find a good agreement with Eq.~(\ref{eq12}). For $E_J/E_C=1$ this is illustrated in Fig.~\ref{R_beta} by the curves proportional to $N^{-(2\alpha-2)}$. A similar agreement was found for $E_J/E_C=2$.

In this Letter we presented a new algorithm for the simulation of resistively shunted Josephson junctions and similar physical systems. The algorithm combines local updates in frequency space with rejection-free cluster updates which exploit the symmetries of the cosine potential. The cluster moves dramatically improve the sampling efficiency, allowing us to study junctions at temperatures which are 100 times lower than what has previously been possible. This has enabled us to reliably confirm conjectured properties of the Schmid transition.

This algorithm will be essential in the study of Josephson junction arrays. Already for the two-junction model with charge relaxation \cite{Refael1} an interesting phase is predicted to occur, which exhibits superconducting phase coherence between the leads, while the individual junctions are insulating.
At intermediate values of $E_J/E_C$ the quantum phase transition might be controlled by an intermediate coupling fixed point with as yet unknown properties. For extended systems, such as chains and 2D arrays, floating phases and lines of fixed points with continuously varying exponents have been predicted \cite{Tewari, Refael2}. Our algorithm will enable the numerical investigation of these models.

We acknowledge support by the Swiss National Science Foundation and helpful discussions with H. G. Evertz, Ch. Helm and G. Refael. This work was started at the Kavli Institute for Theoretical Physics, UCSB. The calculations have been performed on the Asgard Beowulf cluster at ETH Z\"urich, using the ALPS library \cite{ALPS}.

\end{document}